\documentclass[fleqn]{article}
\usepackage{graphicx}
\usepackage[figuresright]{rotating}
\usepackage{axodraw}
\usepackage[T1]{fontenc}
\begin{document} 

%
{\centering \Large {\bf
A Data Acquisition and Monitoring System
for the Detector Development Group at FZJ/ZEA-2}
}

\vspace{1.0cm}  
{\centering \large 

Riccardo Fabbri$^{a}$\\

\vspace{0.3cm}      
{\large 
  $^a$ Forschungszentrum Juelich (FZJ), J\"{u}lich \\
  E-mail: r.fabbri@fz-juelich.de\\
}}

  \vspace{1.5cm}
  \hspace{3.7cm} 
{\centering \large 
  \today
}

\vspace{3.5cm}
\begin{abstract}
  The central institute of electronics (ZEA-2) in the Forschungszentrum 
  J\"{u}lich (FZJ) has developed the novel readout electronics 
  JUDIDT~\cite{JUDIDT} to cope with high-rate data acquisition 
  of the KWS-1 and KWS-2 detectors in the experimental Hall at the 
  Forschungsreacktor M\"{u}nchen FRM-II~\cite{FRM-II} in Garching, 
  M\"{u}nchen. 
  
  This electronics has been then modified and used also for 
  the data-acquisition of a prototype for an ANGER Camera~\cite{ANGER} 
  proposed 
  for the planned European Spallation Source. To commission the electronics, 
  software for the data acquisition and the data monitoring has been 
  developed. In this report the software is described. 

\end{abstract}

{\large \vspace{-18cm} \hspace{-4cm} 
  Forschungszentrum J\"{u}lich\\

  \hspace{-4.cm}
  Internal Report No. XXX
}

\newpage  

\tableofcontents
  \vspace{-20cm} \hspace{6cm} 
\newpage  

\section{Introduction}
%
This note describes how to install and to use the DAQ program suite
developed for the Neutron and Gamma Detector Group (Arbeit Gruppe Neutron und
Gamma Detektoren, AGNuGD) at ZEA-2 in FZJ. 
The documentation also provides a description
of how the code is structured with the scope to give to the future developers
the possibility to easily implement modifications and further developments.

The program was initially developed for the commissioning of the 
JUDIDT2~\cite{JUDIDT} electronics, designed and manufactured at ZEA-2,
to be used as a readout system for an ANGER Camera prototype within 
the XX Collaboration. During the
commissioning its functionality was extended to perform spectrum analysis of
plastic scintillators by using the ACQIRIS digitizers.
The description here given refers to the
software version V2.0. Later changes and implementations should not invalidate
the main procedures here shown.

\section{Installation of the DAQ Software}
%
The suite is distributed as a zip archive DAQ\_Vxx.zip (the tag xx referring to
the version number, as 0.1, 0.2, and so on). After copying the zip file to the
wanted location, the archive can be unzipped (under Linux man can use the
command ``unzip <zip\_file>'') and the directory DAQ\_Vxx will be there
unpacked. In that root directory the sub-directories of each of the DAQ
programs are found, and they contain the source code for their build:

\begin{enumerate}
   \item {\bf LIB\_JUDIDT} \\                       
         ==> library of the JUDIDT electronics; 
   \item {\bf DAQ\_LIB} \\                            
         ==> library of the DAQ;
   \item {\bf DAQ\_SERVER} \\                   
         ==> the DAQ Server;
   \item {\bf DAQ\_CLIENT} \\                 
         ==> the DAQ Client in terminal mode and the libSocket;
   \item {\bf DAQ\_CLIENT\_GUI} \\           
         ==> a graphical interface to communicate with the SERVER;
   \item {\bf ONLINE\_MONITOR} \\     
         ==> the graphical monitor to watch the data online 
             (or the data previously taken);
   \item {\bf RUNLOG} \\                
         ==> the GUI to store in an external file information on the runs.
\end{enumerate}

These programs above listed are the main constituents of the DAQ suite, and
are described in the sections 4.x.

Additional folders and files are present in the unzipped folder. One group is
needed to build the software, and it is foreseen to provide the necessary
flexibility to be platform independent. Tools to compile with gcc, nmake,
Visual Studio and the QT IDE are given. In principle, the QT environment
should provide a cross-platform framework, being its configuration files 
(.pro files) configured to compile in both Linux and Windows systems. By 
using QT
functions many calls are already cross-platform; for example, from the Client
GUI external programs (as the DAQ server or the Online Monitor) can
be started.  
The reference build system is QT, while the other frameworks are periodically
but not regularly updated, and their maintenence is not regularly ensured. 

The compilation folders and scripts are the following:

\begin{enumerate}
   \item {\bf Makefile.gcc} \\
         ==> the script to compile the packages with gcc
   \item {\bf Makefile.nmake}\\
         ==> the script to compile the packages with nmake 
   \item {\bf QT\_CREATOR} \\
         ==> the folder with the configuration files (*.pro) 
         to compile within the QT IDE
   \item {\bf VS2008}\\
         ==> the folder with the configuration files (*.vcproj) to compile 
         within Visual Studio
   \item {\bf bin} \\
         ==> where the executables are locally stored
   \item {\bf include} \\
         ==> include files common to more packages
   \item {\bf lib} \\ 
         ==> libraries locally stored
   \item {\bf CONFIGURE} \\
         ==> Configuration files are here stored:
   \begin{itemize}
         \item {\bf configure.sh} \\
               ==> to perform the initial configuration in Linux 
         \item {\bf configure.bat} \\
               ==> to perform the initial configuration in Windows \ .
   \end{itemize}
   \end{enumerate}

As first step the user has to run the configure.sh(.bat)
script in order to create some relevant folders. Under Linux the installation
of the libraries should require root privileges. During the configuration
the following three directories are created, according to which operating 
system is running (Linux, XP, or the CYGWIN environment for XP):
\begin{enumerate}
   \item Where the binary files of the accumulated data are stored
   \begin{itemize}
          \item {\bf Linux:} \\             
                 ==>   /scratch/DAQ\_REPOSITORY
          \item {\bf XP:} \\
                 ==>  c:{\textbackslash}DAQ\_REPOSITORY
          \item {\bf XP/CYGWIN:}\\
                 ==> /scratch/DAQ\_REPOSITORY 
   \end{itemize}
   \item Where the executables will actually run, and where the program
         configuration files are located
   \begin{itemize}
          \item {\bf Linux:}\\
                ==> /scratch/DAQ\_WORKING\_DIR
          \item {\bf XP:}\\
                ==> c:{\textbackslash}DAQ\_WORKING\_DIR
          \item {\bf XP/CYGWIN:}  \\
                ==> /scratch/DAQ\_WORKING\_DIR
   \end{itemize}
   \item Where the executables are globally located
   \begin{itemize}
          \item {\bf Linux:} \\
                ==> /home/<user\_home\_directory>/bin
          \item {\bf XP:}\\
                ==> c:{\textbackslash}bin 
          \item {\bf XP/CYGWIN:}\\
                ==> /home/<user\_home\_directory>/bin \ .
   \end{itemize}
\end{enumerate}

Note that to compile the drivers for the JUDIDT2 readout system the driver,
library and include files for the Plx tools should be present in the
system. The Plx tools are needed to control the FPGA in the SYS interface and
in the JUDIDT2 module. See the JDAQ.pro file in the
QT\_CREATOR/LIB\_JUDIDT for clarifications. The same applies for the ACQIRIS
system. In addition, also the libraries and the includes of 
ROOT~\cite{ROOT} developed at CERN are
needed, because the GUI packages are built against those libraries, exploiting
their long and mature development to fulfill the requirements of the analysis
in several fields of physics. Please note that when using external libraries
as the ROOT ones, the same compiler should be used to correct identify the 
symbols of the used objects saved in the library.

When no error appears during the compilation, then all the packages
should have been located in the installation directory, which was setup during
the initial configuration procedure. 

Additional folders and files are present:
\begin{enumerate}
    \item {\bf DOCUMENTS} \\
           ==> contains this and additional documentation;
    \item {\bf README}\\
          ==> A short description about how to compile all packages
\end{enumerate}

\section{Overview of the DAQ Programs}
\label{sec:Overiew}
\begin{figure}[h!]
    \hspace{-2.3cm}
      \includegraphics[height=12cm,width=16.cm]{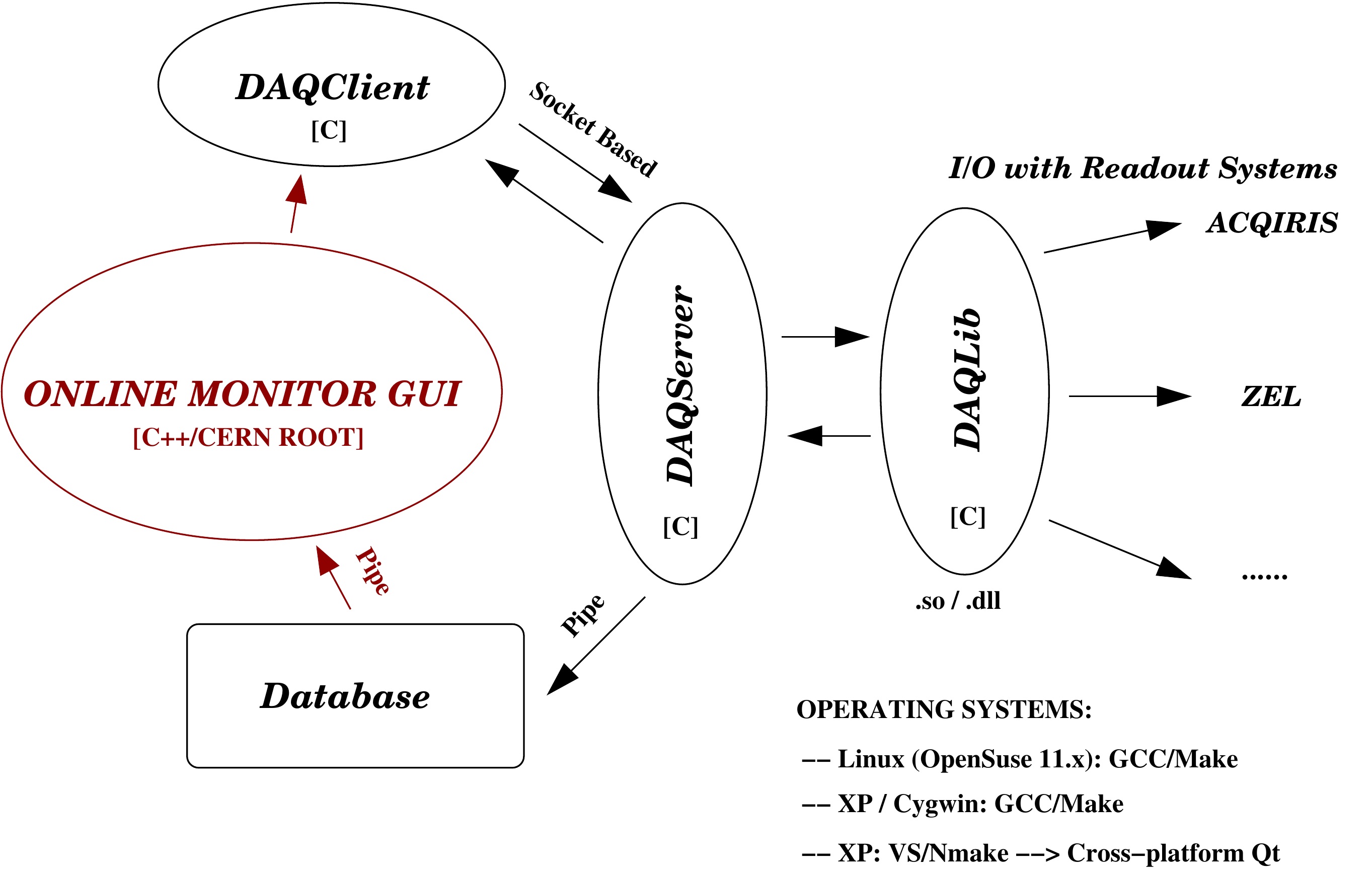}
  \caption{Structure of the client-server architecture of the DAQ.}
  \label{fig:DAQ_structure}
\end{figure}

This DAQ programs are structured in a client-server  
framework, as shown in Fig.~\ref{fig:DAQ_structure}.
The server (DAQServer) runs listening to clients (DAQClient), which 
send it commands via a socket~\cite{SOCKETS}. 
It is eventually the user-level interface 
to a shared library which interacts with the low-level driver 
of the selected readout system.  
%

As soon as the DAQ starts, the data are stored in binary format
into a file saved in the DAQ\_REPOSITORY directory. The data 
can be online or offline inspected 
by the online monitoring
GUI (OnlineMonitor), based on the ROOT libraries. 

In principle all the DAQ operations can 
be done in terminal mode, in cases when the amount of 
computer resources (CPU and memory) could be an issue. 
A friendly GUI (DAQClientGUI), based on ROOT, 
has been also developed to control, directly from its menu, 
the data acquisition and the online monitoring.

The modular structure of this DAQ suite provides 
many advantages with respect to having a single block
of software. It allows in fact the implementation of 
changes to single parts, without altering the remaining 
components. As an example, the OnlineMonitor could be implemented
using graphical libraries different from ROOT, e.g. 
using the LabView interface, or additional features of 
the program can be written without the need of modifying 
the DAQ server and clients. 

On the other side, when a fast monitoring of the data 
(e.g. at acquisition rates as fast as the Megahertz) is needed 
then possibly this solution appears to be not the best, being 
the data acquisition delayed by the continuous i/o to the
database (for example a file).

\section{DAQ Server}
\begin{figure}[b!]
  \hspace{-1cm}
      \includegraphics[height=8cm,width=14.cm]{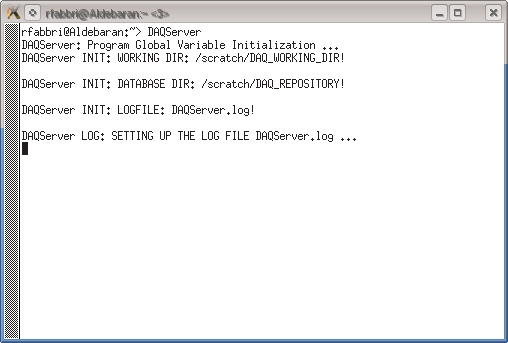}
  \caption{The server is here started directly from a terminal.}
  \label{fig:DAQ_SERVER}
\end{figure}

%
The server (DAQServer) is a terminal mode program, 
which can be started either directly from a 
terminal (Fig.~\ref{fig:DAQ_SERVER}), or using 
the main user GUI. Note that this last option 
also opens a terminal. 
%

As soon as the server is started, a LOG
file is opened in the DAQ\_WORKING\_DIR, and  every 
activity executed is recorded there. The LOG file can 
be online accessed under Linux via 
'tail -f DAQ\_WORKING\_DIR/DAQServer.log'.

The server then opens a socket at the port 5540 (this 
number is for the moment hardcoded), and starts to 
listen at commands sent by any number of independent 
clients.

When a command is received, either to configure 
the hardware or the data acquisition, the server 
calls the relevant function in the DAQ library. 
Here, according to the used readout system 
the corresponding proper function is used; at the server 
side every command disregards the underlining hardware;
it is the DAQ library which takes care to use the 
proper driver with respect to the readout system.

\section{DAQ Client}
%
The client (DAQClient) is a simple program which takes 
as command line argument an option and, when needed, its value. 
Via a socket the command is sent to the server which, 
after processing it, sends back the answer to the 
received command, which could be also a
value of the hardware configuration. This communication
is done using the DAQSocket library.  

\begin{figure}[t!]
   \hspace{-3cm}
      \includegraphics[height=12cm,width=18.cm]{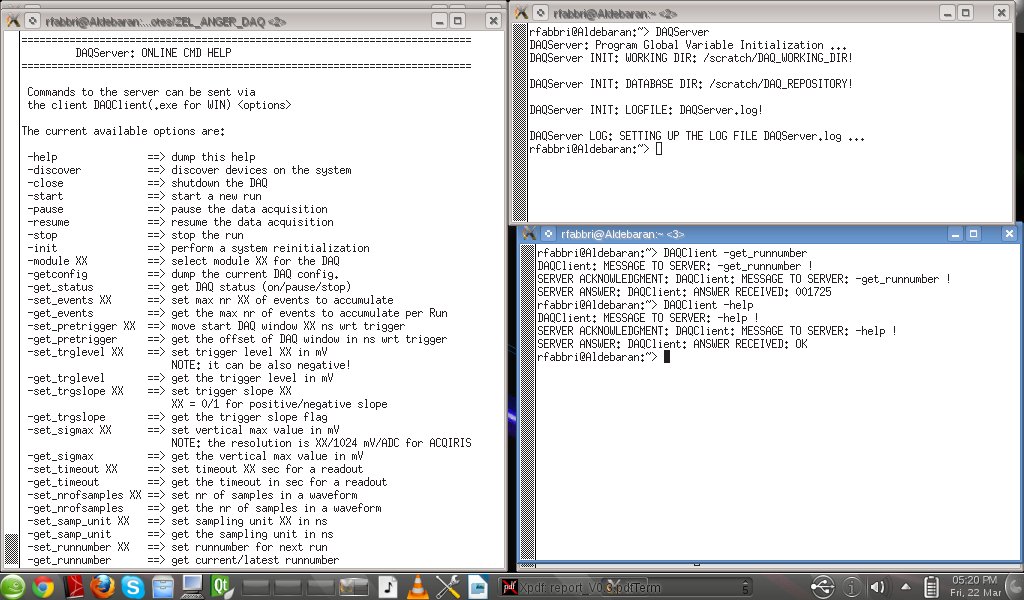}
  \caption{Example of sending a command to the server. The 
           user via a client asks the server for the list of 
           the available options,
           which is dumped into the LOG file by the server.}
  \label{fig:SERVER_CLIENT_EXAMPLE}
\end{figure}

In case no option is sent an error is dumped to 
the terminal. To know which commands are available
the option '-help' should be used (see the 
bottom right terminal in the screenshot shown in 
Fig.~\ref{fig:SERVER_CLIENT_EXAMPLE}), 
and a list of options will be dumped to the server LOG 
file (leftmost displayed terminal in the picture).

After receiving the answer from the server the client always exits.
Commands can be sent to the server by, in principle, 'infinite'    
clients, each one running in an independent terminal. This is 
possible because what is important here to communicate with the 
server, is the existence of a socket at a specific port. 

To start (stop) the data acquisition, the user should simply 
sending the command '-start' (-stop) to the serve. The default 
run number is taken to be the largest already stored run number, 
increased by one unit.   

\section{The DAQ Socket Library}
%
As previously mentioned, the communication between 
server and clients is based on a library 
(libSocket.so/libSocket.dll) which uses a socket for exchanging 
information between them. The working 
principle of a socket is based on the transmission
of character strings via virtual channels (i/o ports 
for the operating system). A shared library function 
with three arguments is used for this purpose. 
The library arguments are pointers to strings.  

A command is transmitted to the server giving the socket 
library the pointer to the command string as its first argument. 
The second and third arguments are the pointers to 
additional strings.
The library acknowledges to have received a certain command
using the string which is pointed by the second argument
of the library. It sends then the command string to the server 
via the socket, and  it remains active waiting for the answer 
from the server via the opened socket. The received string 
is copied to the allocated memory pointed by the third 
argument of the library. 
The client can now retrieve the server answer 
accessing this last address.
The memory needed for these strings is allocated by the client 
before issuing a command.

\section{The Main User Frontend: DAQClientGui}
\begin{figure}[t!]
   \hspace{-3cm}
      \includegraphics[height=12cm,width=18.cm]{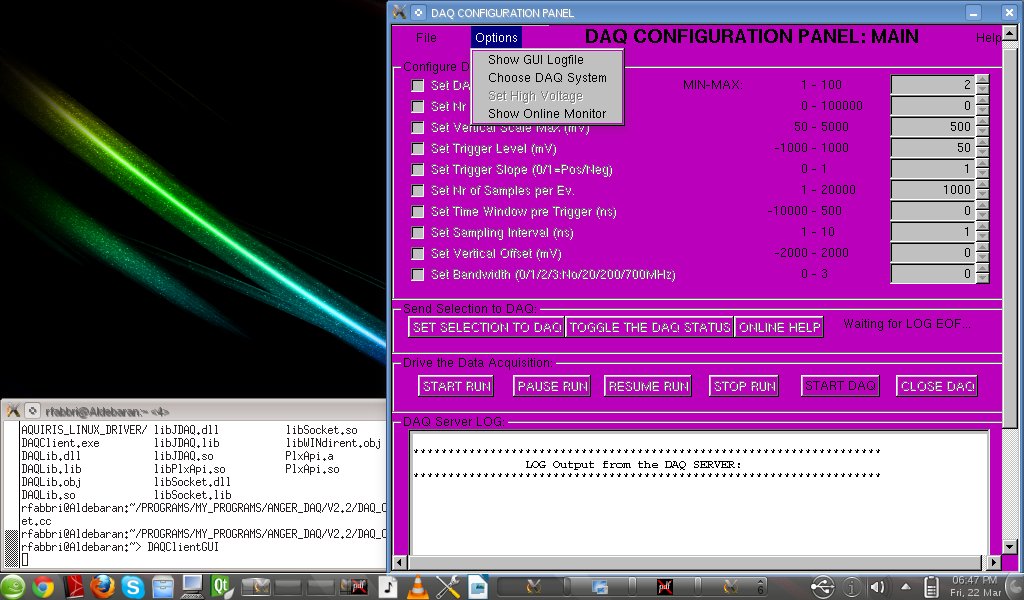}
  \caption{Directly from the GUI the DAQ can 
           be started and closed, the readout system can be chosen
           and the OnlineMonitor program can be launched from there.}
  \label{fig:DAQ_CLIENT_GUI}
\end{figure}

%
A graphical interface (based on the ROOT libraries) was 
developed to help the user to access all the main components 
of the DAQ the server. As shown in the screenshot of 
Fig~\ref{fig:DAQ_CLIENT_GUI}, directly from the GUI the DAQ can 
be started and closed, the readout system can be chosen
(at the moment only the JUDIDT and the ACQIRIS systems are implemented), 
and the OnlineMonitor program can be launched from there.

Once a readout system is set (via the option 'Choose DAQ System' in the 
menu) then the DAQ setting can be configured by activating the 
relevant buttons and choosing the corresponding values in the main 
panel. The selection done can be set to the hardware clicking the 
'SET SELECTION TO DAQ' button. The status of the DAQ can be always 
be accessed using the button 'TOGGLE THE DAQ STATUS' or scrolling 
the output in the server LOG file displayed in the 'DAQ Server LOG'
panel. Note that the DAQ configuration can be performed only 
when data acquisition is not running. This is because all the data
in a run should be accumulated with the same configuration, which is
saved in the header of the run file.

The LOG activity of this GUI can be accessed in a separate canvas via 
the menu option 'Show GUI Logfile'. The Help option gives the user the
possibility to open this documentation in the Acrobat Reader, and the  
History option displays the progress done during the several versions 
of the program.

\section{The DAQ Library}
%
The DAQ library is possibly the core of this data 
acquisition software.
It is a collection of functions to control the 
hardware, and according to the selected readout electronics
its relevant low-level driver function is called. 

When the data acquisition is started, and the data are indeed
accumulated, the value for each event in each channel of the 
hardware is stored in a global structure, whose memory location
is returned back to the server at every readout. This will then 
save the data to an external binary file, clearing the memory 
allocated by the library, and starting a new hardware readout.  
The default configuration keeps on accumulating the data also 
when a run is terminated, increasing automatically the run number 
of one unit.

Differently from the ACQIRIS system, which allows to store 
entire waveforms, 
the JUDIDT electronics provides already the peaking amplitude
(calculated by the internal FPGA). The same data structure 
is maintained by saving the JUDIDT data as single-point 
waveforms.

In principle, following what done using two systems, 
the library can be expanded to accommodate additional readout 
system, keeping the high-level software (Server, Client and 
OnlineMonitor) almost unaffected by the performed changes.
The activity done by the library functions is documented in the
server LOG, whose memory address is given to the library 
functions by the server.

\section{The Online Monitor GUI}
\begin{figure}[b!]
   \hspace{-1.5cm} 
      \includegraphics[trim=0cm 11.5cm 0cm 0cm, clip=true,
                       height=11.cm,width=15.cm]{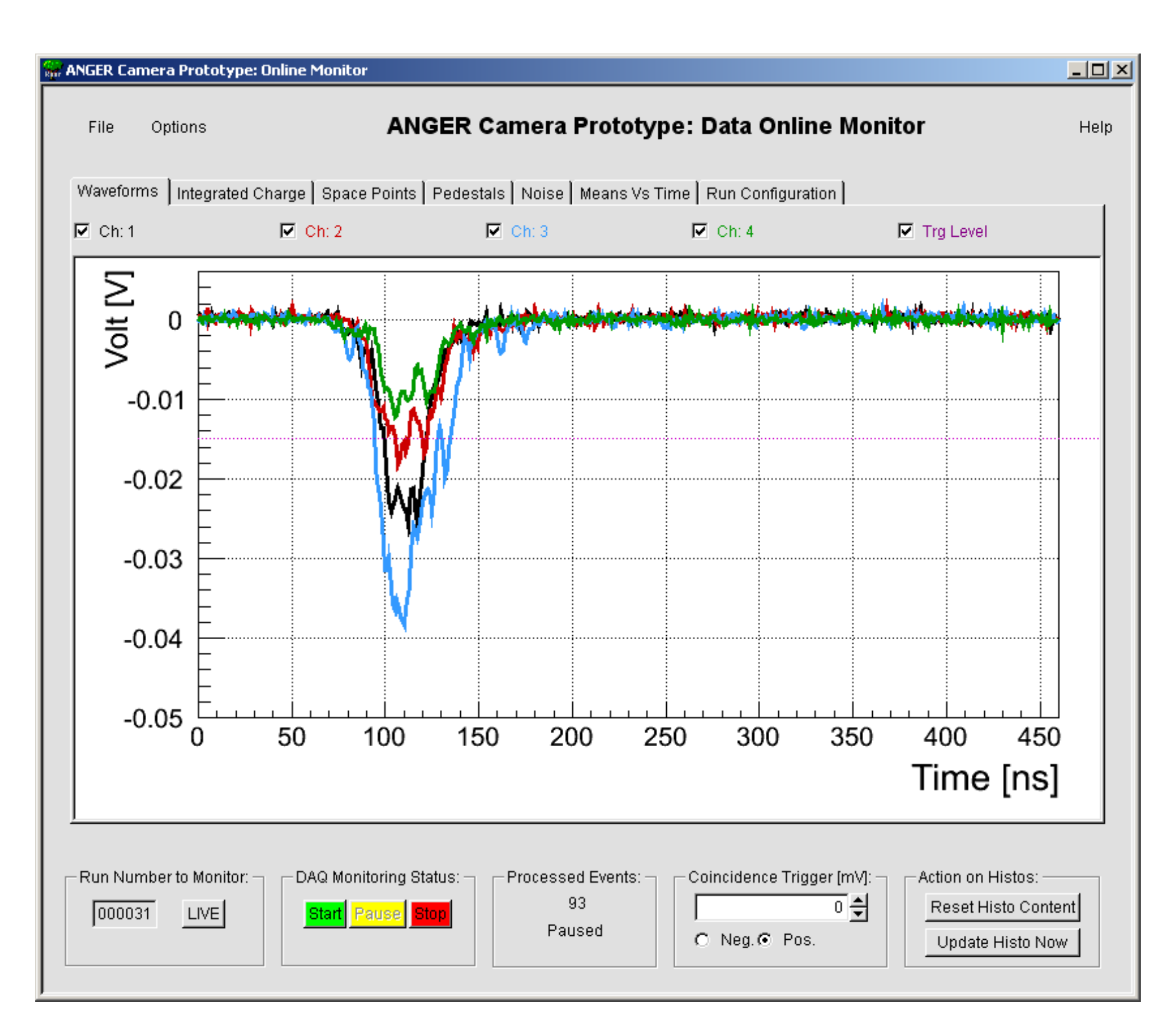}
   \caption{The accumulated data can be monitored online and offline
           directly from a GUI interface. In this example the 
           coincidence of signals from four photomultipliers, originating 
           from a neutron transit in a gas-filled Anger Camera is shown.
           The photomultipliers are here not yet gain-matched.}
   \label{fig:ONLINE_MONITOR}
\end{figure}

%
The friendly graphical user interface OnlineMonitor, based on ROOT,
has been developed to control online the quality of the
data provided by the data acquisition system, Fig.~\ref{fig:ONLINE_MONITOR}.

With the option 'Live' active the program monitors the data 
of the ongoing run (after retrieving the current run number from the
server). Previously accumulated data can be analyzed by providing the
corresponding run number. An error will be prompted whenever the 
data file will not be found. In the 'Live' mode new runs are 
automatically monitored without any action from the user.

At startup the program reads the steering file 
\mbox {\bf ONLINE\_MONITOR.conf}
to give some global variables a value different from the default 
hard-coded one. The format is the following: <var>\%<Value>.
At the moment only the choice of the data repository directory 
is implemented, although in principle additional variables could be 
there implemented.  

The header of the run is shown in the 'Run Configuration' Tab, as soon as the 
run is processed, displaying the following slowcontrol parameters:\\ 
\begin{tabular}{|l|l|}
    \hline
    \hline
    Run Number                       & Readout System, \\
    Start Time of the run            & Number of readout channels,\\
    Number of sampling for waveform  & Sampling time unit,\\
    Delay time                       & Max amplitude accessed\\
    ADC resolution                    & Pedestal offset\\
    Trigger level (during the DAQ)   & Trigger slope \\
    Line impedance & \\
    \hline
    \hline
 \end{tabular}\\ \\ 

Additional tabs show, for each input line, the integrated signal charge, 
the calculated pedestal and noise and the corresponding averaged value vs 
time average (as well as the rates and the temperature vs time; 
the temperature is at the moment not readout, and therefore not displayed 
yet).

During the monitoring the user can choose, as a first rough analysis 
of the data, to provide a common threshold to all the input channels,
either positive or negative. 

For documentation purposes, via the 'File' menu entry, a screenshot 
of the entire program window, or the histograms in the displayed 
canvas can be saved. The format of the generated plot is forseen to 
be chosen in the 'Options' menu entry (not yet implemented). At the 
moment the default format is the postscript.

The tab 'Integrated Charge' shows the calculated total charge integrated 
by the software in case of the AQUIRIS readout which saves entire 
waveforms. In case of the JUDIDT system the integration is performed
inside the electronic (in this case one measurement point is given 
every single event readout).

With the ACQIRIS system at the moment the pedestal is calculated 
also in presence of a neutron signal considering the entire 
waveform; here the value with the maximum number of counts is
considered as pedestal mean. Around this value a small range 
of signal value is used to generated a distribution whose RMS 
is considered as system noise during this event. In order to 
have this technique reliable some data before the risetime 
of the signal should be accumulated.

In case of the JUDIDT system, a 'Calibration' flag should be 
given (via the menu in the DAQClientGUI). Then the data 
are accumulated via a software trigger and could in principle
be considered, e.g. being out of the beam, as system pedestal 
events. This distribution, expected to be Gaussian, is shown in the 
'Pedestals' Tab.

The mean and RMS for the data collected every 
minute are used for the calculation of the time dependence of 
the pedestal and noise in each input line, and are presented in the 
'Means vs Time' Tab. 
Also, the RMS calculated as above mentioned is used to fill the Noise 
histogram presented for every readout channel in the 'Noise' Tab.  

At the moment the Tabs 'Integrated Charge', 'Pedestal' and 'Noise'
allows to show four channels within the available number of channels 
in the analyzed run. The possibility to choose the rate for the
histogram refresh (which can be time consuming) in those three Tabs
and also in the 'Space Points' Tab is given by th an entry widget: at 
high DAQ rates a large value is suggested to keep the monitoring updated
with the data acquisition. On the other side, in case of weak 
sources a  large refresh value could keep the user waiting too long 
for a refresh. Please note that a refresh can be always forced by clicking
on the 'Update Histo Now' button in the bottom Status Bar of the GUI.
In case of the ACQIRIS system the refresh rate does not affect the 
waveform display, which is always performed every sequential waveform.
In the Status section of the GUI  is also present the option 
'Reset Histo Content'; in this case the 
data sofar accumulated in all histograms will be deleted, except for 
the graphs shown in the 'Means vs Time' Tab. 

An additional canvas can be shown to present the LOG entries 
generated by the program. To show or hide it the menu option 
'Show Logfile' can be used. 

In order to correctly process the data content in the binary file
the updated version of the include file DataStructure.h 
should be used at the compilation time.
This issue originates from unwanted but necessary modifications 
in the run header motivated by the data analysis needs. 
The old runs have been modified offline considering the 
latest implementations of the data stream structure. 

\section{The Data Format}
\begin{figure}[b!]
   \hspace{-1.5cm} 
      \includegraphics[height=4.cm,width=15.cm]{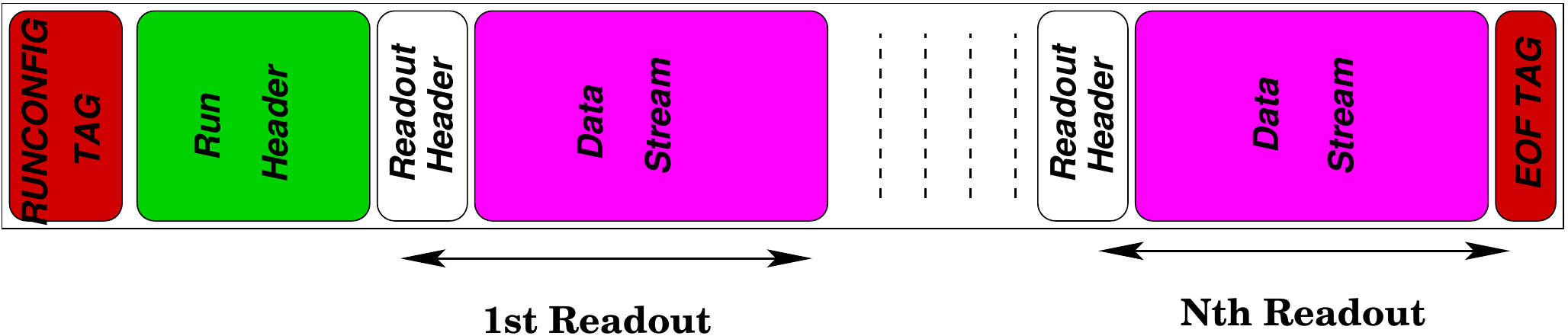}
   \caption{Diagram describing the structure of the streamed data saved in 
            the binary file.}
   \label{fig:DATA_STRUCTURE}
\end{figure}

%
The data are saved in a compressed binary format following the scheme 
shown in Fig.~\ref{fig:DATA_STRUCTURE}.
The first information which should be saved is the run configuration
introduced by the TAG {\bf \#RUN\_CONFIG}, and can be interpreted via the 
C-structure RUN\_CONF:\\\\
%
typedef struct \{\\

\hspace{0cm} \begin{tabular}{lll}
     unsigned long int & RUN\_NUMBER;          & // Run Number            \\
     unsigned long int & RUNSTART\_TIMESTAMP;  & // Start of the Run      \\
     char   & READOUT\_SYSTEM[8];              & // Which Readout System  \\
     int    & NR\_OF\_CHANS;                   & // Nr of Channels        \\
     int    & NR\_OF\_SAMPLES;                 & // Samples in each Event \\
     int    & TRIGGER\_SLOPE;                  & // Trigger Slope         \\
     int    & COUPLING\_FLAG;                  & // Impedance FLAG        \\
     double & DELAY\_TIME;                     & // DAQ Window pre-trigger (secs)\\
     double & SAMPLING\_INTERVAL;              & // Sampling Unit (secs)  \\ 
     double & TRIGGER\_LEVEL\_LOW;             & // Trigger Level         \\ 
     double & Offset;                          & // Volt Offset   \\
     double & FullScale;                       & // Volt Scale    \\\\
\end{tabular}
\} RUN\_CONF;\\

A C-call like\\\\

\hspace{0.5cm} "fread( \&RUN\_CONF, sizeof( RUN\_CONF ), 1, POINTER\_TO\_FILE);"\\\\
should allow to fill all the variables declared in the RUN\_CONF structure.
It is clear that not all variables will obtain a non null value, according 
to the readout system (e.g. the JUDIDT system in this DAQ system does not 
save the waveform but only the peaking amplitude, therefore in this case 
the number of samples per event is one).
At this stage all variable objects dynamically allocated which depend on the 
number of readout channels (e.g. signal, pedestal and noise histograms)
are deleted and reallocated according to the new size provided. This operation
is performed only once at the beginning of the run. 

Soon after the run header should appear the header of the first data readout, 
and the call \\

\hspace{0.5cm} \mbox{"fread( \&DataFlag, sizeof( DataFlag ), 1, POINTER\_TO\_FILE );"}\\\\
will fill the C-structure for the data header:\\\\
typedef struct \{\\

\hspace{0cm} \begin{tabular}{lll}
     int               & DAQ\_TRIGGER;         & // Type of data (Normal/Calib)          \\
     int               & NR\_OF\_READOUT;      & // Readout Counter        \\
     unsigned long int & TIMESTAMP;            & // Start of the Readout   \\
     int               & NR\_OF\_EVENTS;       & // Events in each channel  \\
     long int          & SIZE\_OF\_STREAM;     & // Size of Data Stream    \\\\
\end{tabular}\\
\} DATA\_FLAG;\\

After retrieving how many events per channel are stored in a particular
readout, the analyzer can read the amplitude value of each event following 
this "for loop" sequence:\\\\
for\_loop\_over\_channels \{

   for\_loop\_over\_events\_in\_channel \{
 
      \hspace{0.5cm} for\_loop\_over\_sampling\_in\_event \{ // only one with JUDIDT

      \hspace{1.5cm} ....

      \hspace{0.5cm} \}   //  Loop over samplings in one event

   \}      \hspace{0.5cm} //  Loop over events in one channel\\
\}         \hspace{1.0cm} //  Loop over channels

Knowing how to easily read the data from the binary files
the analyzer can, in principle, perform also a more detailed 
analysis on the data with his own code; it is enough to have
a single include file.

\section{The Run LOG}
\begin{figure}[b!]
   \hspace{-1.5cm} 
      \includegraphics[trim=0cm 11.5cm 0cm 0cm, clip=true,
                       height=14.cm,width=16.cm]{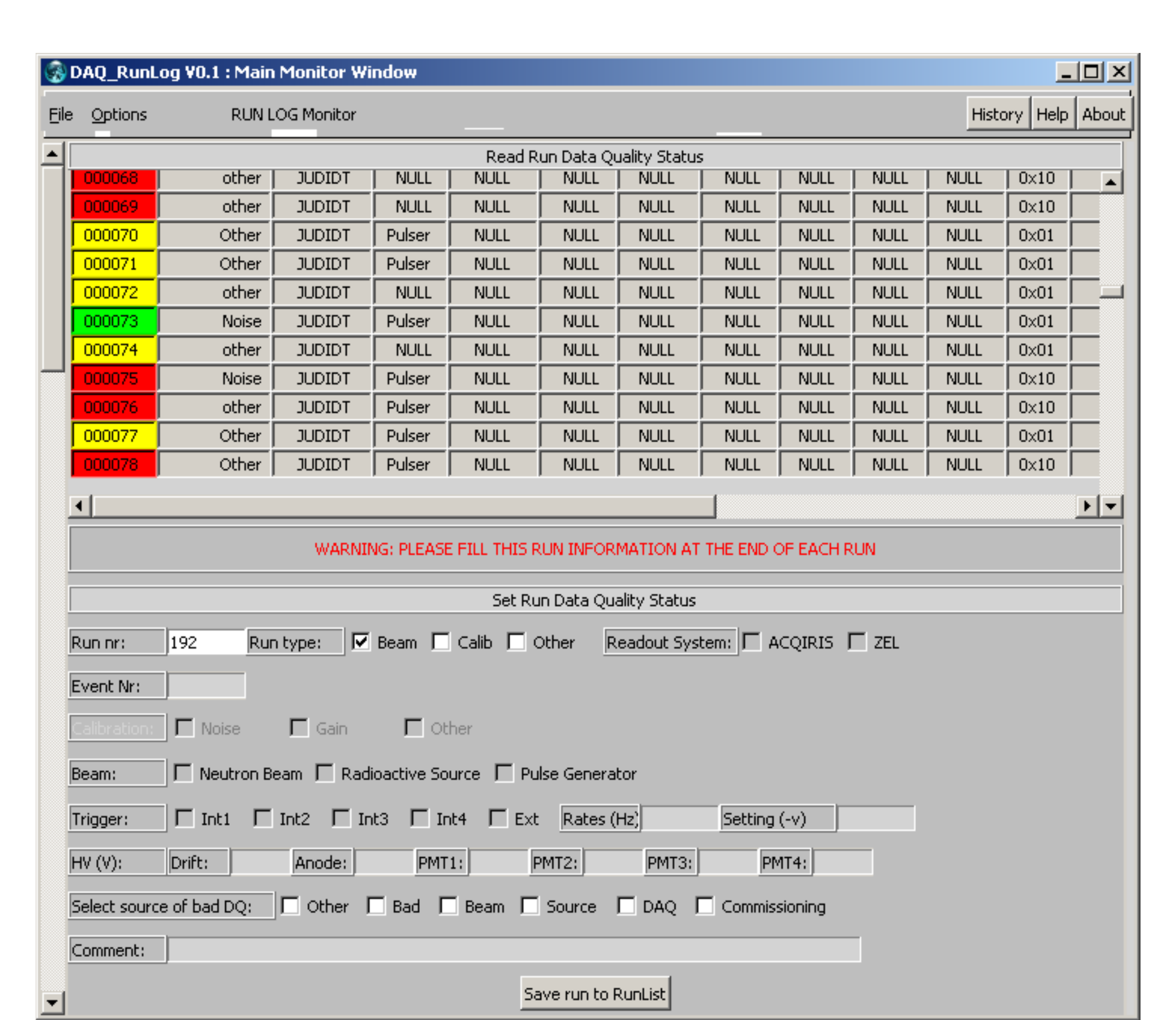}
   \caption{Screenshot of the GUI DAQ\_RunLog.tcl.}
   \label{fig:RUN_LOG}
\end{figure}

%
A small GUI has been developed in Tcl/Tk (due to the 
flexibility and simplicity of this interpreted language)
to store during the data acquisition some 
information relevant to the ongoing run. A screenshot of the 
program is presented in Fig.~\ref{fig:RUN_LOG}.

At the end of each run the user can set some relevant parameters
for the accumulated data, which are then stored in the ascii file
RunList.dat when typing "Save run to RunList". Then a new empty 
form will be presented to the user with the run number increased 
of one unity.

In case an error is made, there is a menu option to retrieve the 
previously saved 
settings for that specific run, and to change them. In the menu there
is also the possibility to delete the information of the last 
saved run, that is of the last record in the ascii file.

Each record of the file refers to a specific run, and 
contains the values of all the relevant variables designed 
for a specific project (in this case, the commissioning of
the JUDIDT electronics). It is clear that, while the needs 
of a project might change, different variables could be implemented.  

According to the selected run type, Normal (Beam), Calib or Other,  
a color code marks the cell of the run number. Normal data taking 
is marked with green; otherwise the yellow code is used. In case a test 
is made, or the accumulated data are for whatever reason bad and useless, the 
source of bad quality should be checked in the section 
"Select source of bad DQ". The run will then be marked with red color.

Some parameters can be set only if they are relevant to the 
specific type of ongoing run. As an example, if normal data taking 
is proceeding the calibration section remains inactive, and will 
be saved in the ascii file as NULL.  

The most striking advantage to use such a program is the 
possibility to easily and fast select subsets of data to analyze 
without going through all the binary data files. 

Although, in the opinion of the author, this interpreted language 
is really optimal for this type of applications (easy and fast to implement),
nevertheless the possibility to convert the program in C/ROOT
is considered in future versions.
This choice is driven by the advantage to use only one language 
in the maintenance of the entire package suite.

\section{The JUDIDT Library}
%
This library was developed in the ZEL-2 
institute, and embedded and tuned in a DAQ code for 
an Online Monitor based on LabView, thus in a
monolithic software needed for a specific project. 

In principle, being developed by third parties it 
should not be mentioned here; nevertheless, for its 
use in this separate  DAQ project some modifications were needed, 
to keep only the driver for the input/output 
with the electronics, and to decouple it from the
LabView components for the monitoring. 

Additional modifications were needed to make the
functions of this driver as much as possible consistent
with the driver for the ACQIRIS readout, also 
implemented in this DAQ. 

The modified code can be compiled in the DAQ\_CLIENT folder
in combination with the DAQClient with the compilers gcc or nmake,
or even better using the QT development environment 
in QT\_CREATOR/LIB\_JUDIDT. 

\section{Conclusions}
%
We have presented in this note a new software for the 
data acquisition based on the Server-Client architecture, 
and based on the socket concept for send and receive commands. 
All the components of the package were described 
in separate sections.

It is clear that the code is at the moment not really 
a general purpose code, being in many parts tuned to 
the forseen project to investigate the JUDIDT electronics.
Nevertheless the code has been designed to be made of 
separate independent components, and  make strong use of 
pointers (thus of 
dynamically allocated memory arrays with variable size). 
This aspect should allow an easier implementation of additional 
features, if needed in the future.
 
The publication of the results of the analysis performed on 
the JUDIDT readout system, also in combination with the 
Anger Camera prototype is on going~\cite{ANALYSIS}.

%
\begin{quotation}
      \vspace{0.5cm}
  \begin{center}
      {\bf Acknowledgments}
      \vspace{0.5cm}
  \end{center}
 The author gratefully acknowledges 
 U.\ Clemens, R.\ Engels, and G.\ Kemmerling  
 for their valuable technical contribution and 
 suggestions to the work here presented. 

\end{quotation}



\end{document}